# Dissipative cnoidal waves (Turing rolls) and the soliton limit in microring resonators

Zhen Qi[1,*], Shaokang Wang[1], José Jaramillo-Villegas[2], Minghao Qi[3], Andrew M. Weiner[3], Giuseppe D'Aguanno[1], Thomas F. Carruthers[1], and Curtis R. Menyuk[1]

[1]*University of Maryland at Baltimore County, 1000 Hilltop Circle, Baltimore, MD 21250, USA*
[2]*Technological University of Pereira, Cra. 27 10-02, Pereira, Risaralda 660003, Colombia*
[3]*Purdue University, 610 Purdue Mall, West Lafayette, IN 47907, USA*
*\*Corresponding author: zhenqi1@umbc.edu*

**Single solitons are a special limit of more general waveforms commonly referred to as cnoidal waves or Turing rolls. We theoretically and computationally investigate the stability and accessibility of cnoidal waves in microresonators. We show that they are robust and, in contrast to single solitons, can be easily and deterministically accessed in most cases. Their bandwidth can be comparable to single solitons, in which limit they are effectively a periodic train of solitons and correspond to a frequency comb with increased power. We comprehensively explore the three-dimensional parameter space that consists of detuning, pump amplitude, and mode circumference in order to determine where stable solutions exist. To carry out this task, we use a unique set of computational tools based on dynamical system theory that allow us to rapidly and accurately determine the stable region for each cnoidal wave periodicity and to find the instability mechanisms and their time scales. Finally, we focus on the soliton limit, and we show that the stable region for single solitons almost completely overlaps the stable region for both continuous waves and several higher-periodicity cnoidal waves that are effectively multiple soliton trains. This result explains in part why it is difficult to access single solitons deterministically.**
© 2019 Optical Society of America under the terms of the OSA Open Access Publishing Agreement

***OCIS codes:*** (190.4380) Nonlinear optics, four-wave mixing; (140.4780) Optical resonators; (190.3270) Kerr effect.

http://dx.doi.org/10.1364/osa.XX.XXXXXX

## 1. INTRODUCTION

It has long been known that single solitons are a special limit of a more general family of stationary nonlinear waves that are almost universally referred to as cnoidal waves in the plasma and fluid physics communities [1–6]. When loss and gain can be neglected, many fluid, plasma, and optical systems can be described at lowest-order by the nonlinear Schrödinger equation (NLSE). Examples in optics include light propagation in optical fibers [7], passively modelocked lasers [8], and microresonators [9–11]. In the NLSE approximation, solitons can be analytically expressed in terms of hyperbolic-secant functions [sech $(x)$], while more generally cnoidal waves can be expressed in terms of Jacobi elliptic functions [cn $(x)$, sn $(x)$, dn $(x)$] [12]. When loss and gain are present—as is the case in all experimental fluid and optical systems—then exact analytical solutions to the equations that describe these systems almost never exist. Nonetheless, it is common to refer to pulse and periodic solutions as "solitons" and "cnoidal waves" respectively. In the past 15 years, it has become increasingly common to refer to real-world solitons as "dissipative solitons" [13], and by analogy it seems reasonable to refer to real-world stationary periodic solutions as "dissipative cnoidal waves."

The literature on cnoidal waves in fluids is extensive, but that is not the case in optics, except in the soliton limit. (See,

however, [14–16] for theoretical discussions.) Solitons or cnoidal waves that are described at lowest order by the NLSE are created by a balance between nonlinearity and dispersion. The pulses in optical fiber propagation experiments are typically separated by many times their duration in order to avoid instabilities [17]. Even in experiments in optical fiber ring resonators [18] or passively modelocked lasers [19] with many pulses in the cavity, it is useful to model the pulses as a periodic train of weakly interacting solitons, rather than as a special limit of cnoidal waves. However, the small dimensions of microresonators make it easier to observe more general cnoidal waves than is possible in other optical systems, and they have frequently been observed [20–27], although often remarked upon only as a stepping stone on the path to single solitons [20–23].

In recent years, a large experimental effort has focused on obtaining broadband microresonator combs using single solitons, and this effort has achieved remarkable experimental successes, including the demonstration of octave-spanning combs [28, 29], a $2f$-$3f$ self-referenced comb [30], and an optical frequency synthesizer [31]. However, these combs cannot be obtained in a straightforward way. In contrast to many modelocked lasers, the microresonators do not "self-start," i.e., when the microresonator is turned on, single solitons do not simply appear. Approaches to generate single solitons include pump tuning, thermal control, and engineered spatial mode interactions [32–35]. However, there is often a random element in the generation of single solitons [34, 35], and the search for robust, deterministic paths to obtain broadband combs continues. Recent work indicates that is possible to generate single solitons deterministically by using trigger pulses [36], by using two pumps [37–40], or by adding an additional continuous wave component [41]. However, these approaches make the system significantly more complex. Another drawback to the use of single bright solitons to generate combs is that they make inefficient use of the pump. Typically, less than 1% of the pump power is transferred to the comb [23].

In this paper, we will show that more general cnoidal waves can potentially solve these problems. Cnoidal waves can be broadband, while at the same time they can be straightforwardly accessed and use the pump more efficiently. They are sufficiently easy to obtain that they have frequently been observed, although, as previously noted, they have often been passed over. Their robustness has, however, been previously noted [24, 25].

An essential difficulty in studying the potential of cnoidal waves is that there is a large parameter space to explore. As previously noted, single solitons correspond to period-1 cnoidal waves in which the azimuthal repetition period of the light in the microresonator cavity is equal to the entire azimuthal period, so that there is only a single pulse in the cavity. By contrast, cnoidal waves can have any periodicity, and a period-$N_{\text{per}}$ cnoidal wave has an optical amplitude that repeats $N_{\text{per}}$ times in one azimuthal period. Moreover, the cnoidal waves with a given periodicity $N_{\text{per}}$ can vary greatly from structures in which the light intensity is never close to zero to structures that are effectively a periodic train of solitons or a soliton crystal. Indeed, this variability is their great advantage. We will show that it is possible to continuously tune the microresonator parameters to effectively obtain a train of solitons without ever passing through a chaotic regime, as is required to obtain single solitons.

Given the size of the parameter space to be explored, more theoretical guidance is needed, and direct solution of the evolution equations is inadequate for reasons that we will discuss shortly.

To address this issue, we have adapted a unique set of computational tools that were previously developed to study lasers [42–45] and that allow us to rapidly determine where in the parameter space cnoidal wave solutions exist and to find their frequency spectra. These tools are based on dynamical systems theory [46, 47] and complement widely available theoretical tools that are based on direct solution of the evolution equations. The basic idea is that once a cnoidal wave solution of a given periodicity has been found, one varies parameters, finding the cnoidal wave solutions as parameters vary by solving a nonlinear root-finding problem, instead of solving the evolution equations. In parallel, one determines the stability of the solution by finding the eigenvalues of the linearized equations. Once a stability boundary is encountered, where the cnoidal waves become unstable or cease to exist, we track the boundary [42]. This approach is computationally rapid, and it allows us to identify all the stable cnoidal wave solutions that exist for a given set of parameters, in contrast to standard evolutionary simulations that will typically only identify a single solution, depending randomly on the initial conditions that are used. This approach also allows us to find unstable as well as stable cnoidal waves. That is important because unstable cnoidal waves can persist for a long time, and this approach allows us to calculate that time scale. When the lifetime of an unstable cnoidal wave is long, standard evolutionary simulations can incorrectly conclude that it will be stable.

The basic ideas that we are using are old, dating back at least to Maxwell's study of the stability of Saturn's rings [46]. Soliton perturbation theory is based on these ideas, and they have been adapted by Haus [47] and others [48] to study passively modelocked lasers. They have been used by Matsko and Maleki [49] to study soliton timing jitter and by Godey et al. [50] to study the stability of continuous waves in microresonators. Barashenkov et al. [51, 52] have applied these ideas to the study of multiple soliton solutions of the driven-damped NLSE [53], which in the microresonator community is typically referred to as the Lugiato-Lefever equation (LLE) [54]. However, almost all this work is based on analytical approximations of the stationary solutions. (The work in [51, 52] is an exception.) In our approach, we calculate computationally exact stationary solutions, which allows us to accurately study the solutions in any parameter regime. That is particularly important when studying cnoidal waves in the anomalous dispersion regime, which is the focus of this paper. While analytical solutions exist for the cnoidal wave solutions of the LLE when damping is neglected [55], these solutions are disconnected from the stable cnoidal wave solutions in the anomalous dispersion regime except when the periodicity is small, as discussed in Sec. 4 and Sec. 1 of Supplement 1. As a consequence, the utility of a perturbation theory based on these analytical solutions appears likely to have limited value.

The microresonator community has developed a nomenclature that differs in some respects from the nomenclature in common use in the nonlinear waves community. The use of "Lugiato-Lefever equation" instead of "driven-damped nonlinear Schrödinger equation" is one example. The periodic solutions of the LLE have been referred to as "Turing rolls," "primary combs," and "soliton crystals," as well as "cnoidal waves" or "dissipative cnoidal waves." The choice "Turing roll" connects the cnoidal wave solutions with earlier work on pattern formation [56]; the choice "soliton crystal" emphasizes the particle-like nature of solitons and connects the periodic solutions to single soliton solutions. As noted previously, our choice of "cnoidal waves" connects these solutions to the large body of work on periodic solutions in fluids, plasmas, and nonlinear waves.

The remainder of this paper is organized as follows: In Sec. 2, we present our basic equations, discuss our normalizations and how they relate to the experimental literature, and review our dynamical approach. In Sec. 3, we present our results on the stability of the cnoidal wave solutions—showing where in the parameter space stable solutions exist. We discuss their accessibility and how to optimize their parameters, including their bandwidth. In Sec. 4, we discuss the single soliton solutions. Our approach sheds new light on the difficulties in accessing them. Finally, Sec. 5 contains the conclusions.

## 2. BASIC EQUATIONS, NORMALIZATIONS, AND DYNAMICAL METHODS

Light propagation in microresonators is described at lowest order by the LLE [9–11], which may be written

$$T_R \frac{\partial A}{\partial \tau} = -\frac{i\beta_2}{2}\frac{\partial^2 A}{\partial \theta^2} + i\gamma |A|^2 A + \left(-i\sigma - \frac{l}{2}\right) A + i\sqrt{P_{\text{in}}}, \quad (1)$$

where $\tau$ is time, $\theta$ is the azimuthal coordinate, $A$ is the field envelope of a single transverse optical mode, $T_R$ is the round-trip propagation time and is the inverse of the free-spectral range (FSR), $\beta_2$ is the dispersion coefficient, $\gamma$ is the Kerr coefficient, $\sigma$ is the pump detuning, $l$ is the damping coefficient, and $P_{\text{in}}$ is the input pump power that is in the cavity mode that pumps the comb. The dispersion coefficient $\beta_2$ is related to the physical dispersion $\beta_{\text{ph}}$ by the relation $\beta_2 = \left(n^3 R/T_R^2\right)\beta_{\text{ph}}$, where $R = v_g T_R/(2\pi)$ is the mode radius and $v_g$ is the group velocity.

Equation (1) can be written in terms of three normalized parameters. One standard way to normalize Eq. (1) is to let $t = l\tau/2T_R$, $\psi = (2\gamma/l)^{1/2}A$, $\beta = 2\beta_2/l$, $\alpha = 2\sigma/l$, and $F = i(8\gamma/l^3)^{1/2}\sqrt{P_{\text{in}}}$, in which case Eq. (1) becomes

$$\frac{\partial \psi}{\partial t} = -\frac{i\beta}{2}\frac{\partial^2 \psi}{\partial \theta^2} + i|\psi|^2 \psi - (i\alpha + 1)\psi + F. \quad (2)$$

We will be focusing in this paper on the anomalous dispersion regime where $\beta < 0$. In this case, it is useful to let $x = (2/|\beta|)^{1/2}\theta$, in which case Eq. (2) becomes

$$\frac{\partial \psi}{\partial t} = i\frac{\partial^2 \psi}{\partial x^2} + i|\psi|^2 \psi - (i\alpha + 1)\psi + F. \quad (3)$$

Only the parameters $\alpha$ and $F$ appear explicitly in Eq. (3). The third parameter appears implicitly in the boundary conditions. The domain of the azimuthal coordinate, $\theta = [-\pi, \pi]$, is replaced by the domain $x = [-L/2, L/2]$, where $L = (8\pi^2/|\beta|)^{1/2}$ is the mode circumference $2\pi R$ normalized to the dispersive scale length. The solution obeys periodic boundary conditions, i.e., $\psi(-L/2, t) = \psi(L/2, t)$. This normalization is useful because $|\beta| \ll 1$ for realistic experimental parameters, and it allows us to conveniently explore the limit $L \to \infty$.

An alternative is to normalize Eq. (1) with respect to the detuning, assumed positive, so that $t = \sigma\tau/T_R$, $x = (2\sigma/|\beta_2|)^{1/2}\theta$, $\psi = (\gamma/2\sigma)^{1/2}A$, $\delta = l/2\sigma$, and $h = (\gamma/2\sigma^3)^{1/2}\sqrt{P_{\text{in}}}$. In this case, Eq. (1) becomes

$$\frac{\partial \psi}{\partial t} = i\frac{\partial^2 \psi}{\partial x^2} + 2i|\psi|^2\psi - (i + \delta)\psi + ih = 0. \quad (4)$$

The periodic domain is now given by $x = [-L_\delta/2, L_\delta/2]$, where $L_\delta = (8\pi^2\sigma/|\beta_2|)^{1/2}$. This normalization is theoretically useful since it can be used to conveniently connect the computational solutions of the LLE to the analytical solutions in the limit $\delta \to 0$ [55]. However, it is not useful for describing experiments, in which the detuning and the pump power are varied, and the detuning can have any sign. In the body of this paper, we therefore use the normalization in Eq. (3). In Sec. 1 of Supplement 1, we use the normalization of Eq. (4) to show that the analytical cnoidal wave solutions with $\delta = 0$ are disconnected from the stable cnoidal wave solutions with $\delta \neq 0$ unless the periodicity of the cnoidal wave solutions ($N_{\text{per}}$) is small. Cnoidal waves are wavetrains that repeat periodically within the mode circumference $\theta = [-\pi, \pi]$ or $x = [-L/2, L/2]$.

In experimental work, typical values of $F$ range from 0 to 4, while typical values of $\alpha$ range from $-5$ to 10. The corresponding device parameters vary greatly, depending on the material from which the device is made and its size. Table 1 shows the experimental parameters that correspond to $\alpha = 1$ and $F = 1.5$ for several different experimental devices. The normalized mode circumference $L$ ranges between 30 and 200.

The standard evolutionary methods for solving the LLE are typically based on the split-step method and its variants [7]. We use a variant that we recently demonstrated is significantly faster computationally than most other variants when large loss and gain are present [58]. In a standard evolutionary simulation, one starts from an initial condition, and one steps the LLE forward in time, sometimes allowing the equation's parameters to also change as a function of time. When a stationary solution is observed at the end of the simulation, one has learned where in the parameter space that particular stationary solution can be accessed, starting from the initial conditions and along the path in the parameter space that were used in the simulation. By contrast, dynamical methods for solving the LLE reveal where in the parameter space the stable solutions exist, but do not reveal how to access them. The dynamical methods function much like an x-ray machine that tells a surgeon where a bone is broken, but not how best to repair it. Hence, the evolutionary and dynamical methods are complementary.

While evolutionary methods are widely used in the microresonator community, dynamical methods are not. Our implementation is an adaptation of computational methods that were developed to study passively modelocked lasers [42–45]. Our starting point is a cnoidal wave solution that we find using evolutionary methods with equation parameters for which the solution is highly stable. This solution is stationary and satisfies Eq. (3) with its time derivative set equal to zero,

$$0 = i\frac{\partial^2 \psi}{\partial x^2} + i|\psi|^2 \psi - (i\alpha + 1)\psi + F. \quad (5)$$

We now solve Eq. (5), while allowing $F$ or $\alpha$ to vary, which becomes a nonlinear root-finding problem. We use the Levenberg-Marquart algorithm [59], which is a variant of the well-known Newton's method.

In parallel with solving Eq. (5) to find the cnoidal wave solutions, we determine their stability. We write

$$\psi(x,t) = \psi_0(x) + \Delta\psi(x,t), \qquad \bar{\psi}(x,t) = \psi_0^*(x) + \Delta\bar{\psi}(x,t), \quad (6)$$

where $\Delta\psi$ and $\Delta\bar{\psi}$ are small perturbations of the cnoidal wave solution $\psi_0(x)$ and its complex conjugate $\psi_0^*(x)$. The perturbations $\Delta\psi$ and $\Delta\bar{\psi}$ obey the linearized equation

$$\frac{\partial \Delta\boldsymbol{\Psi}}{\partial t} = \mathcal{L}\Delta\boldsymbol{\Psi}, \quad (7)$$

where

$$\Delta\boldsymbol{\Psi} = \begin{bmatrix} \Delta\psi \\ \Delta\bar{\psi} \end{bmatrix} \quad (8)$$

**Table 1.** Physical device parameters for different experimental devices. The first column gives the paper reference. The second column gives the Kerr coefficient ($\gamma$). The third column gives the pump wavelength ($\lambda_p$). The fourth column gives the quality factor ($Q$). The fifth column gives the chromatic dispersion in physical units ($\beta_{\text{ph}}$). The sixth column gives the detuning corresponding to $\alpha = 1$. The seventh column gives the input power corresponding to $F = 1.5$. The eighth column gives the normalized mode circumference ($L$).

| Ref. | $\gamma$ [W$^{-1}$km$^{-1}$] | $\lambda_p$ [nm] | $Q$ | $\beta_{\text{ph}}$ [ps$^2$km$^{-1}$] | $\Delta\lambda$ [fm] | $P_{\text{in}}$ [mW] | $L$ |
|---|---|---|---|---|---|---|---|
| [23] | 633 | 1551 | $3 \cdot 10^6$ | $-61$ | 334 | 9.0 | 31.3 |
| [57] | 800 | 1562 | $3 \cdot 10^5$ | $-47$ | 2913 | 1060 | 106 |
| [21] | 0.405 | 1553 | $4 \cdot 10^8$ | $-9.4$ | 1.97 | 21.6 | 35.1 |
| [24] | 0.228 | 1552 | $1.7 \cdot 10^9$ | $-3$ | 0.46 | 16.2 | 171 |
| [26] | 9.49 | 1562 | $2.5 \cdot 10^8$ | $-6.3$ | 3.34 | 2.03 | 120 |

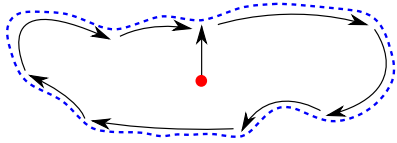

**Fig. 1.** A schematic illustration of the boundary-tracking algorithm. The blue-dashed line indicates the boundary of the stable region. The red circle is the starting point. We obtain the complete stability boundary numerically by moving along the boundary, while tracking its location, as shown with the blue arrows.

and

$$\mathcal{L} = \begin{bmatrix} i\partial^2/\partial x^2 + 2i|\psi_0|^2 - i\alpha - 1 & i\psi_0^2 \\ -i\psi_0^{*2} & -i\partial^2/\partial x^2 - 2i|\psi_0|^2 + i\alpha - 1 \end{bmatrix}. \quad (9)$$

We convert Eq. (7) into an eigenvalue equation, $(\mathcal{L} - \lambda\mathcal{I})\Delta\Psi = 0$, where $\mathcal{I}$ is the identity operator, and we discretize this equation. We then proceed to computationally find the set of all eigenvalues $\{\lambda_j\}$ of this equation. There is always one eigenvalue located at zero, which is due to the translational invariance of Eq. (3) and is immovable. Other than that, all eigenvalues of a stable cnoidal wave solution satisfy $\text{Re}(\lambda_j) < 0$, so that any perturbation except a translation will decay exponentially. All the eigenvalues are real or come in complex conjugate pairs. As we allow $F$ or $\alpha$ to vary, we eventually encounter a stability boundary at which one of three things happen: (1) a single real eigenvalue becomes equal to zero, and the cnoidal wave solution ceases to exist; (2) one or more real eigenvalues pass through zero, and an unstable cnoidal wave solution continues to exist; or (3) the real parts of a complex conjugate pair of eigenvalues become equal to zero, and an unstable cnoidal wave continues to exist. Once a stability boundary is encountered, we track its location, as described in detail by Wang et al. [42] and as shown schematically in Fig. 1. As we track the boundary, we find that transitions occur among these three cases.

The ultimate fate of the system when the parameters pass through a stability boundary must be determined by solving the evolution equations using evolutionary methods. We have found that case (1) leads to a transition from a cnoidal wave solution to a continuous wave solution, case (2) leads to a transition to another cnoidal wave solution, and case (3) leads to a transition to a breather or chaos. These three cases are referred to in the mathematics literature as saddle-node, transcritical, and Hopf bifurcations, respectively [60, 61]. In the mathematics literature, the set of eigenvalues $\{\lambda_j\}$ are referred to as the "spectrum" of the operator $\mathcal{L}$ [62, 63]. Here, we refer to the $\{\lambda_j\}$ as the "dynamical spectrum" in order to avoid confusion with the frequency spectrum. The dynamical spectrum of the LLE has a large amount of symmetry. It is symmetric about the real $\lambda$-axis, since if $\lambda_j$ is an eigenvalue, then so is $\lambda_j^*$. Additionally, it is symmetric about the axis $\text{Re}(\lambda) = -1$. We discuss some features of the dynamical spectrum in Sec. 3 and in Sec. 2 of Supplement 1. While interesting, the details of this structure are of little importance for this study since almost all the eigenvalues correspond to perturbations that rapidly damp out. Our focus is almost exclusively on the eigenvalues whose real parts become equal to zero as the microresonator parameters change.

Once the stable region of a cnoidal wave with a given periodicity has been found, its properties can be rapidly found throughout the region by solving Eq. (5) as $\alpha$ and $F$ vary. We have used this technique to find the fraction of the pump power that goes into the cnoidal wave frequency comb, as well as the comb bandwidth. By solving the evolution equations, we have verified that any solution within the stability boundary can be experimentally accessed in principle by changing the pump power and the detuning.

The dynamical methods that we are using have significant advantages over just using evolutionary methods to identify stable operating regions in the parameter space and paths through the parameter space to access them. First, directly solving the evolution equations for given initial conditions and system parameters can miss stable solutions that exist for those parameters. As shown schematically in Fig. 2, a stable waveform has a basin of attraction in the phase space of all possible waveforms. If the initial conditions are within this basin, then the evolution will eventually converge to this waveform, but, if it does not, then the evolution will not converge to this solution, and it is possible to miss it completely. This issue is particularly important for the cnoidal wave solutions where several different periodicities can be stable at the same point in the parameter space. Second, evolutionary methods yield ambiguous results near a stability boundary since the time for a perturbation to either decay or grow tends toward infinity. Examples of this ambiguity appear in Sec. 3 and in Sec. 1 of Supplement 1. Third, the dynamical methods are computationally rapid, making it possible to determine the stability boundaries for all the cnoidal wave solutions within a broad parameter range. We have typically found that this approach is 3–5 orders of magnitude faster than evolutionary methods, depending on how well one wants to resolve the stability boundary.

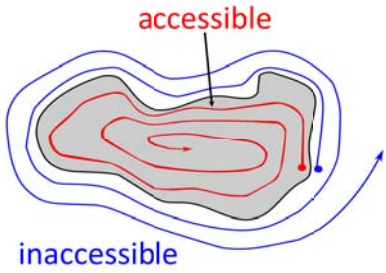

**Fig. 2.** A schematic illustration of a basin of attraction in the phase space that consists of all possible waveforms. If the initial condition, shown as a red dot, starts inside the basin of attraction of a stationary waveform, shown in gray, then the solution will converge to the stationary solution. If the initial condition starts outside the basin, shown as a blue dot, then the solution evolves to another stationary solution or never becomes stationary.

In most computational studies of the dynamics in microresonators, the instabilities are seeded by random numerical noise due to roundoff. This seeding implies a random noise amplitude on the order of $10^{-15}$ in modern-day 64-bit computers. This noise amplitude is many orders of magnitude lower than what is expected in an experimental system, which is largely set by environmental or "technical" noise [64, 65]. By contrast, the randomly varying amplitude due to quantum noise is in the range of $10^{-5}$–$10^{-4}$ for typical parameters as discussed in Sec. 3 of Supplement 1. The growth of the amplitude of an unstable eigenmode is given by $A(t) = A_0 \exp(\lambda_R t)$. As long as $\lambda_R^{-1}$ is small compared to the time scale of the simulation by factors larger than about $\ln 10^{-15} \simeq -35$, then the small level of seeding from roundoff noise does not matter. When making a transition out of a region in the parameter range that is chaotic or where continuous waves are stable into a region where cnoidal waves are stable, the growth of the cnoidal waves is sufficiently rapid that it is not unreasonable to seed the growth from roundoff noise. We will see however that the time scale for instabilities to manifest themselves when making transitions from one cnoidal wave to another can be a sizable fraction of a second in real time, and these instabilities are computationally challenging to simulate. For that reason, we use a quantum noise seed, as described in Sec. 3 of Supplement 1. While the quantum noise power is lower than the technical noise power in microresonators by one to three orders of magnitude [65], it is large enough for instabilities to manifest themselves on physically realistic time scales.

## 3. STABILITY, ACCESSIBILITY, AND OPTIMIZATION OF CNOIDAL WAVES

In Fig. 3, we show the regions where the cnoidal waves are stable for different periodicities $N_{\text{per}}$ in the range $-2 < \alpha < 6$ and for $L = 10, 25$, and $50$. For clarity, we only show some of the stable regions. However, we have found all the stable regions in the parameter range that we show, and a complete map of the stable regions for $L = 50$ is given in Sec. 4 of Supplement 1. When $L \to \infty$, we find that $N_{\text{per}}/L$ tends to a constant for a fixed value of $\alpha$ and $F$, as we show in Fig. 4. The value in the limit $L \to \infty$ is calculated using an approach like that of Godey et al. [50]. More details are in Sec. 5 of Supplement 1. As a consequence, the three-dimensional parameter space in which all solutions of the LLE exist effectively reduces to a two-dimensional parameter space when $L \geq 50$.

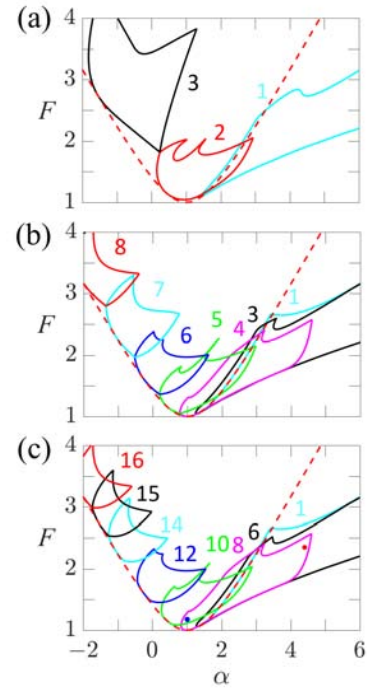

**Fig. 3.** Stable regions of the cnoidal waves for (a) $L = 10$, (b) $L = 25$, and (c) $L = 50$. We show a selection of the stable regions, labeled with $N_{\text{per}}$. The red-dashed curves show the limit below which continuous waves are stable.

In Fig. 3, we see that the region where cnoidal waves are stable form a U-shaped belt in the $\alpha$-$F$ plane. Below this belt, only continuous waves (flat solutions) are stable and are always obtained. Above this belt, there are no stationary solutions; only breathers or chaotic solutions are obtained, depending on the parameters. An analytical expression for this curve is derived in [50] and is given by $F^2 = 1 + (1 - \alpha)^2$. An important transition occurs at $\alpha = 41/30$. Below this value of $\alpha$, stable continuous waves and cnoidal waves do not both exist at the same parameter values. Hence, cnoidal waves can be easily accessed by simply raising the pump power. Above this value of $\alpha$, stable cnoidal waves and continuous waves can both exist at the same parameter values. In particular, continuous waves are stable whenever single solitons are stable. Hence, single solitons cannot be accessed by simply raising the pump power. Additionally, the stable region for single solitons overlaps almost completely the stable region for several higher-periodicity cnoidal waves when $L \geq 25$. The overlap of the stable region for single solitons with other stationary solutions explains in part why they are difficult to access.

In Fig. 5, we show the optical waveform, its frequency spectrum, and its dynamical spectrum for the $N_{\text{per}} = 8$ solution with $L = 50$, $\alpha = 4.4$, and $F = 2.3$, corresponding to the red dot in Fig. 3(c). Figure 5(a) shows the stationary waveform's intensity, $|\psi(x)|^2$. Figure 5(b) shows $P_n$, defined as $P_n = |\tilde{\psi}_n|^2/|\tilde{\psi}_0|^2$, where $\tilde{\psi}_n = (1/L) \int_{-L/2}^{L/2} \psi(x) \exp(-i2\pi n N_{\text{per}} x/L) \, dx$. Each comb line is spaced eight times the FSR apart. This $N_{\text{per}} = 8$ cnoidal wave is effectively a set of eight periodically-spaced

solitons—a periodic train of solitons or a soliton crystal—that is stablized by sitting on a small pedestal. The power $P_n$ of the comb lines falls off exponentially from the central line. This clean, exponential falloff is characteristic of all cnoidal waves and distinguishes them from a random assortment of solitons or other pulses. In Fig. 5(c), we show the dynamical spectrum of the waveform, and in Fig. 5(d), we show an expanded view of a small region near $\lambda = (0,0)$. The dynamical spectrum has a fourfold symmetry about the real axis and the axis $\text{Re}(\lambda) = -1$. There is always an eigenvalue at $\lambda = (0,0)$, corresponding to azimuthal invariance of the solution and hence there is an eigenvalue at $\lambda = (-2,0)$. Except for these two points, we have found that all the eigenvalues on the real axis are degenerate. We see that a group of real eigenvalues cluster near zero. These eigenvalues correspond roughly to modes that shift the amplitudes and phases of the pulses or solitons that make up the cnoidal wave, while eigenvalues along the axis $\text{Re}(\lambda) = -1$ correspond roughly to modes that perturb the pedestal. We explain some of the principal features of this spectrum in Sec. 2 of Supplement 1. However, most of these features are of little relevance for this study. Our focus is almost entirely on eigenvalues that reach the imaginary axis as $\alpha$ and $F$ change, implying that the stationary solution either becomes unstable or disappears.

In Figs. 5(e) and 5(f), we show the optical waveform and its frequency spectrum for $\alpha = 1$ and $F = 1.15$, corresponding to the blue dot in Fig. 3(c). In this case, the $N_{\text{per}} = 8$ cnoidal wave appears to be a modulated continuous wave, rather than a periodic train of solitons. Starting with this $F$ and $\alpha$ and continuously varying the parameters to $\alpha = 4.4$ and $F = 2.3$ makes it possible to lock in place the periodic soliton train in Fig. 5(a).

In Fig. 6, we show examples of the transitions that occur when a stability boundary is crossed, starting in the region where the $N_{\text{per}} = 8$ solutions are stable. In Fig. 6(a), we show the stable regions for $N_{\text{per}} = 8$ and $N_{\text{per}} = 9$, as well as four trajectories in the parameter space. We show the transitions in the dynamical spectrum for all four trajectories, focusing on the eigenvalues that hit or cross the imaginary axis. For clarity, we have omitted other eigenvalues. We show the initial eigenvalues as red dots and the final eigenvalues as blue circles. We also show the stationary solution as a function of the variable that we are varying, $\alpha$ or $F$. In Figs. 6(b) and 6(c), we show the dynamical spectrum and the stationary pulse intensities for trajectory A, for which we start at $\alpha = 2$, $F = 1.29$ and decrease $F$ to 1.27. Slightly below $F = 1.28$, a saddle-node bifurcation occurs, at which point the cnoidal wave solution and hence its dynamical spectrum cease to exist. In this case, the final point in the dynamical spectrum is at $\lambda = (0,0)$. The solution evolves from a cnoidal wave into a continuous wave. In Figs. 6(d) and 6(e), we show the dynamical spectrum and stationary pulse intensities for trajectory B, for which we start at $\alpha = 4$, $F = 2.3$, and we increase $F$. A Hopf bifurcation occurs; the cnoidal wave continues to exist, but it is unstable. It evolves into a breather solution, also referred to a secondary comb. In the Fig. 6(e), we show the region beyond the Hopf bifurcation in gray since no stationary solution exists. In Figs. 6(f) and 6(g), we show the dynamical spectrum and pulse intensities for trajectory C, for which we start at $\alpha = 2$, $F = 1.83$, and increase $F$. A Hopf bifurcation occurs, and the $N_{\text{per}} = 8$ cnoidal wave evolves into a stable $N_{\text{per}} = 9$ cnoidal wave. Finally, in Figs. 6(h) and 6(i), we show the dynamical spectrum and pulse intensities for trajectory D, for which we start at $\alpha = 0.85$, $F = 1.12$, and decrease $\alpha$. A transcritical bifurcation occurs, and the stable $N_{\text{per}} = 8$ cnoidal wave evolves into a stable $N_{\text{per}} = 9$ cnoidal wave. However, the time scale on which this evolution occurs is noticeably longer than is the case for trajectories A, B, and C. That is a common feature of transcritical bifurcations and underscores the importance of using a realistic background noise level and sufficiently long integration times when using evolutionary methods.

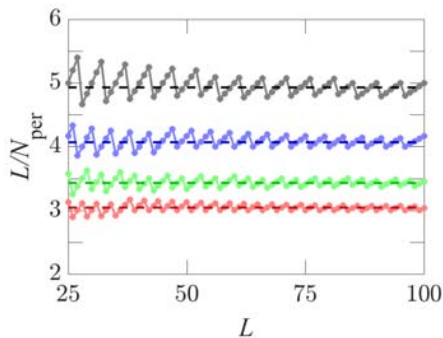

**Fig. 4.** $L/N_{\text{per}}$ vs. $L$ for three values of $(\alpha, F)$: $\alpha = -2$, $F = 3.5$ (red); $\alpha = -1$, $F = 2.6$ (green); $\alpha = 0$, $F = 1.7$ (blue); $\alpha = 1$, $F = 1.2$ (black). Dashed lines show the asymptotic values in the limit $L \to \infty$. We obtained the stationary solution by using an evolutionary approach with a small random initial seed.

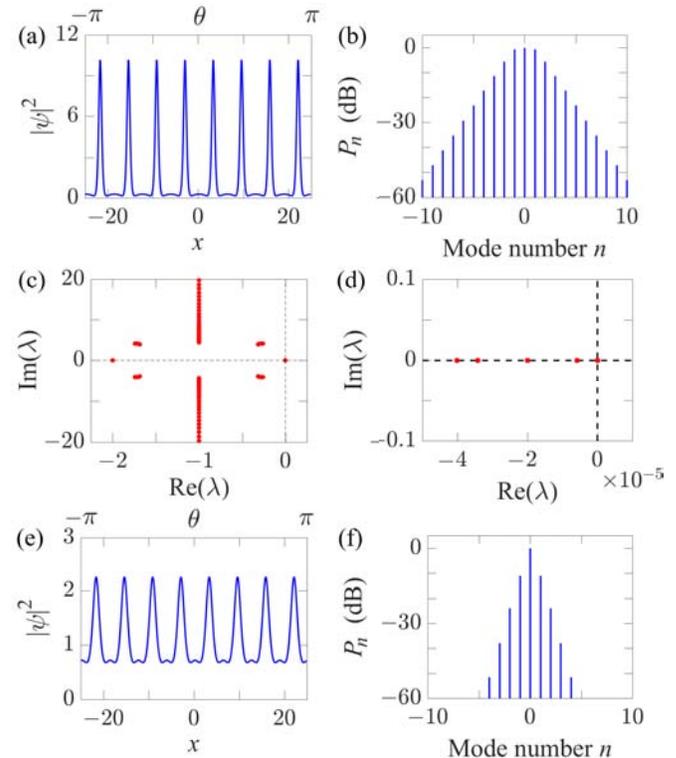

**Fig. 5.** (a) Waveform, (b) frequency spectrum, and (c) dynamical spectrum with $N_{\text{per}} = 8$, $L = 50$, $\alpha = 4.4$, and $F = 2.3$. (d) Expanded view of the dashed rectangular region near $\lambda = (0,0)$ in (c). (e) Waveform and (f) frequency spectrum with $\alpha = 1$ and $F = 1.15$. Figs. 5(b) and (c) are modified from Fig. 2 of [66].

Comparing Figs. 6(d) and 6(f) to Fig. 6(h), we see that $\lambda_R$, the maximum real value that appears in the dynamical spectrum, is about 25 times smaller for a transcritical bifurcation than it is after a Hopf bifurcation for a comparable excursion in the parameter space. As a consequence, the time scale for the instability to manifest itself is considerably longer for a transcritical bifurcation. In Fig. 7, we show the evolution as a function of normalized time and for physical time, using the parameters of Wang et al. [23], starting from the unstable stationary cnoidal wave solutions at the final points of Figs. 6(d), 6(f), and 6(h). As expected the time scale in Fig. 7(c), corresponding to Fig. 6(h) is about 25 times larger than the time scale in Figs. 7(a) and 7(b), corresponding to Figs. 6(d) and 6(f). The dynamical spectrum is particularly useful when studying transcritical bifurcations since without its guidance, one can easily integrate for too short a time to accurately assess the stability of the solution.

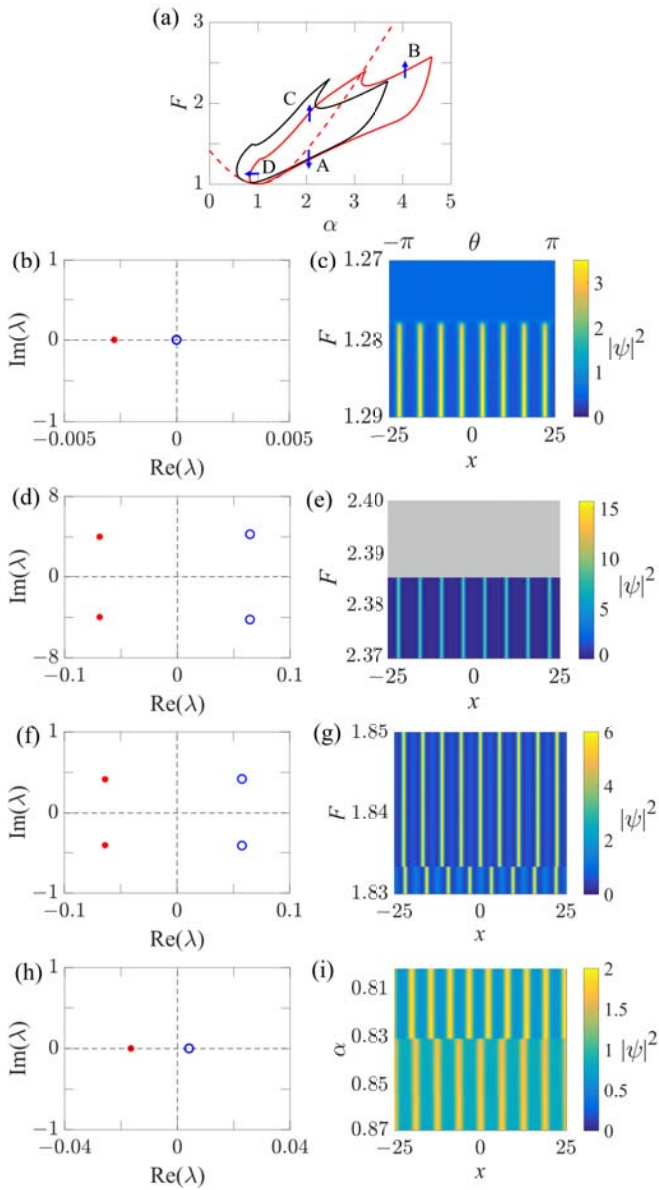

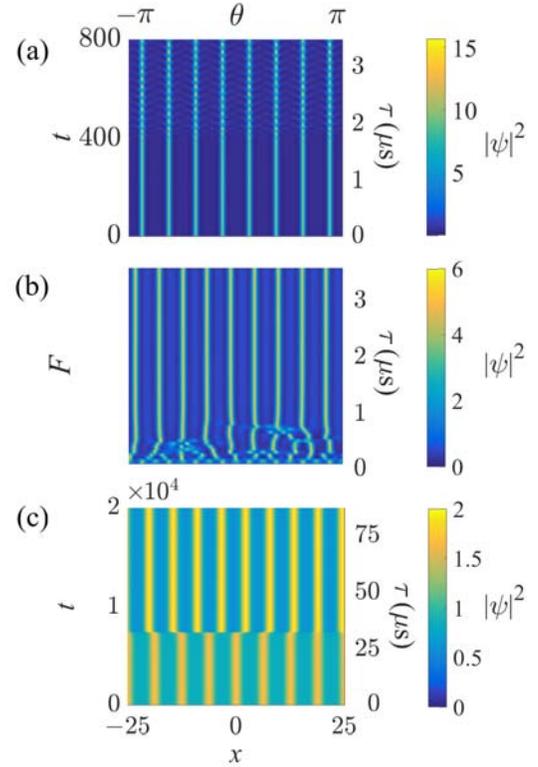

**Fig. 7.** Evolution of pulse intensity corresponding to the final points in trajectories (a) B, (b) C, and (c) D in Fig. 6.

**Fig. 6.** (a) Stable regions for the $N_{\text{per}} = 8$ (black curve) and $N_{\text{per}} = 9$ (red curve) cnoidal waves. The red-dashed curve plots the limit below which continuous waves are stable. A, B, C, and D indicate the four trajectories through the parameter space that we consider. Dynamical spectra and pulse intensities for: (b) and (c) trajectory A, (d) and (e) trajectory B, (f) and (g) trajectory C, and (h) and (i) trajectory D.

In Fig. 8, we compare the stability regions for $L = 50$ with evolutionary simulations for $L \simeq 46$. We show a similar comparison in Sec. 1 of Supplement 1 for the normalization of Eq. (4). In Fig. 8(b), we show the number of peaks in the final solution after jumping from an initial cnoidal wave at $\alpha = 0$ and $F = 6.3$ to another point in the parameter space and remaining at that point for $\tau = 1.8$ $\mu$s [67]. While the results correspond roughly to the stability curves, differences should be noted. The dwell time is not always sufficient to determine the stability, and the location of the stability boundaries is ambiguous. Additionally, the passage time through the parameter space affects the final state. We will discuss this issue in more detail in Sec. 4. By moving along the trajectory in the phase space shown as the solid black curves in Fig. 8(a) or the red-dashed curve in Fig. 8(b), it is possible to stay within the region in the parameter space where the cnoidal waves are stable and deterministically control, with occasional back-tracking, the periodicity of the cnoidal wave that is asymptotically obtained.

In Fig. 9, we show an optimization study for the $N_{\text{per}} = 8$ cnoidal wave when $L = 50$. The point at $\alpha = 4.4$, $F = 2.3$, shown as a red dot in Figs. 3 and 9, corresponds to the frequency spectrum that we show in Fig. 5(b). The point at $F = 1.15$ and $\alpha = 1$ corresponds to the blue dot. The frequency spectrum always decreases exponentially away from the central peak. In Fig. 9(a), we show a contour plot of the rate of exponential decrease, $P_n/P_{n+1}$ (dB) for $n \geq 3$. In Fig. 9(b), we show a contour plot for the ratio $\rho$ of the power in the comb lines to the input pump power, $\rho = (1/F^2) \int_{-L/2}^{L/2} |\psi|^2 (dx/L)$. The corresponding change in the physical ratio of the power in the comb to the power in the pump is given by $\rho_{\text{phys}} = (4/l^2)\rho$, which for the parameters of [23] and a single soliton is given by $\rho_{\text{phys}} = 73.6\%$, which is close to the measured value at the drop port. There is a tradeoff between the efficiency with which the pump is used and the bandwidth of the frequency spectrum. At the point in the parameter space shown as a red dot, we find that $\gamma < 6$ dB, while $\rho \simeq 0.3$. The comb lines in the frequency spectrum are separated by eight times the free spectral range (FSR). The lines at $n = \pm 6$ are down by 30 dB from the central line. For an FSR of 230 GHz, corresponding to the experiments in [23], the corresponding bandwidth is 21 THz, which is a large fraction of the operating frequency of 200 THz. This bandwidth is consistent with the bandwidth of single solitons in microresonators, whose bandwidth has not been enhanced through the use of dispersive radiation. It should be possible to apply techniques that have been used to enhance the bandwidth of single solitons [29, 30] to cnoidal waves with larger periodicity. While the comb line spacing (1 THz) is large, it is consistent with the line spacing of the single soliton comb with large line spacing in the dual-microcomb experiment of [31].

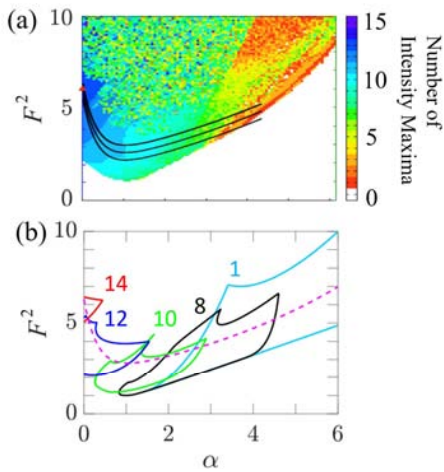

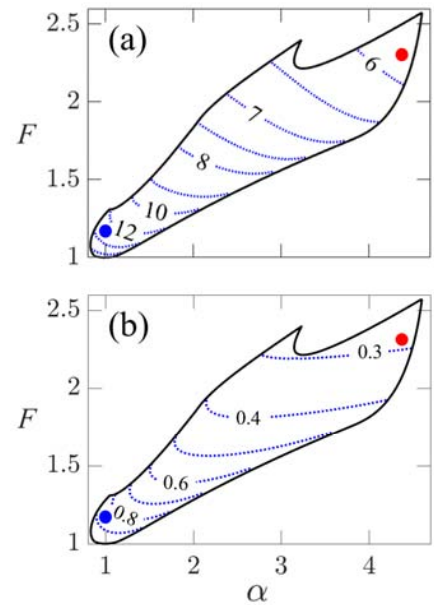

**Fig. 8.** Comparison of the stability map for $L = 50$ with evolutionary simulations from which the stable regions can be inferred. Fig. 8(a) is modified from Fig.2(a) in [67]. The color bar shows the number of intensity maxima in the plot. The solid curves in (a) and the red-dashed curve in (b) show trajectories through the parameter space that never pass through the chaotic region. This figure is modified from Fig. 1 in [66].

**Fig. 9.** Contour plots of (a) the exponential decrease in the frequency spectrum $P_n/P_{n+1}$ (db), $n > 3$, and (b) the ratio of the total power in the comb lines to the input pump power $\rho$ for $N_{\text{per}} = 8$. Fig. 9(a) is modified from Fig. 2(a) in [66].

To characterize the increase in efficiency of pump utilization as $N_{\text{per}}$ increases, we computed $\rho$ for periodicities $N_{\text{per}} = 1$–8 with $\alpha = 4.4$ and $F = 2.3$. We found that $\rho$ increases almost linearly, so that $\rho = 0.038$ when $N_{\text{per}} = 1$ and $\rho = 0.287$ when $N_{\text{per}} = 8$. This linear increase is not surprising since at these values of $\alpha$ and $F$, the cnoidal wave is effectively a periodic train of $N_{\text{per}}$ solitons. Since the total power is nearly proportional to $N_{\text{per}}$ and the comb lines are spaced $N_{\text{per}} \times$ FSR apart, the power in each comb line is nearly proportional to $N_{\text{per}}^2$. That can be important for applications where the power in a single comb line must be made large.

## 4. THE SOLITON LIMIT

Single solitons are a special limit of cnoidal waves for which $N_{\text{per}} = 1$. They share many of the properties of cnoidal waves for which $N_{\text{per}} > 1$. In particular, their frequency spectrum falls off exponentially away from the central peak. Single solitons are also a special limit of another class of nonlinear stationary waveforms, referred to as multi-bound solitons or soliton molecules [52, 68]. Except in special limits where they are single solitons or higher-periodicity cnoidal waves, soliton molecules are characterized by a complicated frequency spectrum.

Single solitons in microresonators are not easy to access and a large experimental effort has been aimed at finding ways to do that [20–26, 28–35, 42]. Here, we focus on understanding the issues with accessing them from a dynamical perspective and we discuss the reasons that there is often a random element in their generation [34, 35]. We extend an earlier discussion that proposed a deterministic path through the parameter space to obtain single solitons [67], and we describe the critical role that is played by the time scale on which one moves through that path.

In Fig. 10, we show the stable regions for the $N_{\text{per}} = 1, 3,$ and 5 cnoidal waves when $L = 50$. For clarity, we do not show the regions of stability for $N_{\text{per}} = 2, 4,$ or 6. However, they may be seen in the map of the stable regions in Sec. 4 of Supplement 1. While the stable region for single solitons ($N_{\text{per}} = 1$ cnoidal waves) is large, it almost entirely overlaps the regions of stability for several higher periodicity cnoidal waves in the parameter range that we show, as well as with continuous waves. Hence, it is not surprising that multiple solitons are often obtained in experiments rather than single solitons. However, the multiple solitons are usually randomly spaced, leading to a complicated frequency spectrum, which is usually undesirable in applications. While these states are unlikely to be stationary states, simulations indicate that they can be long-lived, with lifetimes that far exceed standard simulation runs.

We can obtain insight into this behavior by examining how single solitons are generated. The most common path to obtain single solitons is to start in the blue-detuned region where either continuous waves or high-periodicity cnoidal waves are stable, at the left side of the U-shaped region shown in Fig. 3 [20–

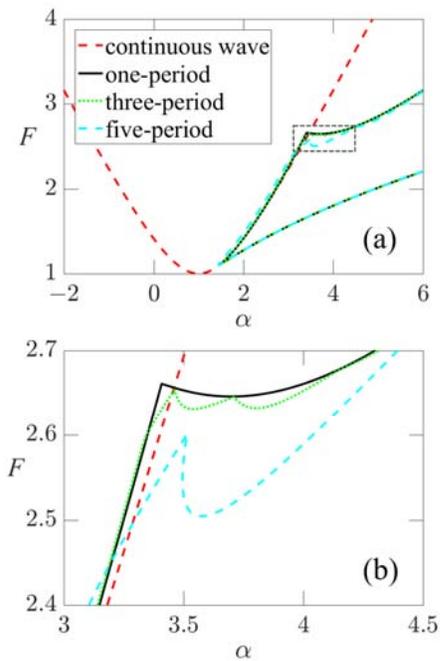

**Fig. 10.** Stable regions for the $N_{\text{per}} = 1$ (black), $N_{\text{per}} = 3$ (green), and $N_{\text{per}} = 5$ (cyan) cnoidal waves. The $N_{\text{per}} = 1$ cnoidal wave is a single bright soliton. Continuous waves are stable below the red-dashed curve. The curves for $N_{\text{per}} = 1, 3,$ and 5 almost completely overlap except in the black dashed rectangular region, of which an expanded view is shown in (b).

25]. Solitons are then obtained by red detuning the pump laser through the chaotic region above the U-shaped region where cnoidal waves are stable into the region where single solitons, other low-periodicity cnoidal waves that are effectively a periodic train of solitons, and continuous waves are stable. This path is labeled A in Fig. 11. An alternative path, labeled B in Fig. 11, is to raise the pump power above the line where continuous waves are stable and then to lower the power until single or multiple solitons are stable [22, 28, 33]. A third alternative, labeled C in Fig. 11, is to tune backward after entering the region where single solitons are stable [34]. Since the soliton duration is inversely proportional to the detuning, backward tuning enhances the interaction between randomly created multiple solitons and thus hastens their collapse into a single soliton, a continuous wave, or another stationary state. In all these paths, the region where single solitons are stable is entered from the chaotic region, in which both temporal and spatial power fluctuations are comparable to the average power and are several orders of magnitude larger than the underlying noise level in the device [64]. As a consequence, the spacing of the solitons when they form will be random and both their number and spacing will vary from shot to shot. Their ultimate evolution will depend on the details of their formation. Since continuous waves and a number of higher-periodicity cnoidal waves are also stable, nothing in principle prevents their formation.

The stable region for single solitons that we show in Fig. 11 is bounded by curves at which either a saddle-node bifurcation occurs (cyan curves) or a Hopf bifurcation occurs (blue curve). In Fig. 12, we show the dynamical spectra for the stable soliton for $L = 50$, $\alpha = 5.5$, and $F = 2.11, 2.45,$ and $2.90$, shown as the red dots in Fig. 11. It is evident that the spectrum dif-fers significantly from the dynamical spectrum that appears in soliton perturbation theory [69] or the Haus modelocking equation [47]. In that case, the only discrete eigenvalues correspond to amplitude, frequency, time, and phase shifts. There are also continuous eigenvalues that correspond to background radiation. In this case, near the value at which the Hopf bifurcation occurs ($F = 2.9$), the least-damped eigenvalues correspond to damped-periodic oscillations of the single soliton. Near the value at which the saddle-node bifurcation occurs ($F = 2.11$), the least-damped eigenvalue corresponds to a damped amplitude and phase modulation. The cluster of eigenvalues near $\lambda = (0,0)$ that was visible in the case of the $N_{\text{per}} = 8$, shown in Fig. 5, and which correspond to damped interactions between the solitons that make up the cnoidal wave, are not present.

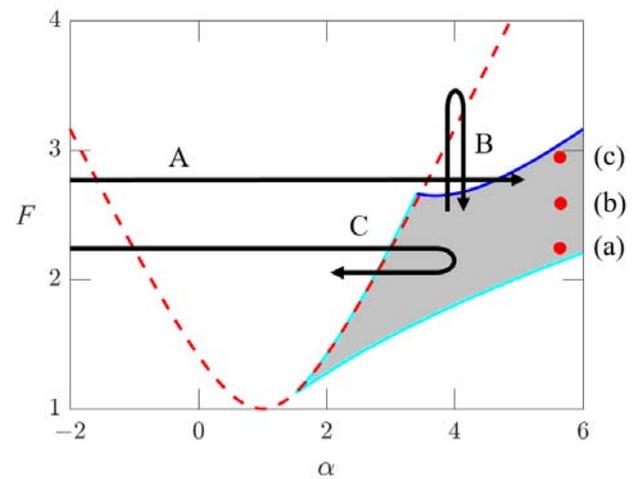

**Fig. 11.** Schematic illustration of three trajectories through the parameter space to obtain single solitons. The gray region indicates where single solitons are stable, and continuous waves are stable below the red-dashed line.

As a consequence, we might expect single solitons to be more robust in the presence of large noise or environmental fluctuations. We have verified, doing a long simulation run for which $t = 10^8$ that the $N_{\text{per}} = 8$ cnoidal wave illustrated in Fig. 5 remains stable. These simulations are computationally taxing, requiring many hours on a standard desktop computer. However, this normalized time $t$ only corresponds to about 1 s of physical time, for a resonator quality factor $Q = 1.67 \times 10^6$, corresponding to the parameters of [67]. Determining their nonlinear stability in the presence of noise fluctuations over the lifetime of an experiment remains to be done.

A promising approach to deterministically obtain low-periodicity, large-bandwidth cnoidal waves, including single solitons, is to avoid the chaotic region [67]. Rather than just changing the detuning or the power, it is advantageous to change both together so that the system moves along one of the paths shown as the black curves in Fig. 8(a) or the red-dashed curve in Fig. 8(b). In this case, transcritical bifurcations occur along the trajectory through the parameter space when cnoidal waves become unstable. The time scale of these instabilities, which is typically milliseconds, is long compared to most simulations, although short compared to most experiments. It is possible to take advantage of this long time scale to deterministically create single solitons [67].

In Fig. 13, we show the the evolution when the system pa-rameters are shifted from $\alpha = 0$ to $\alpha = 6$ along the red-dashed curve in Fig. 8(b) at different speeds and the system is then allowed to evolve to a stationary final state. We set $L = 50$ and use $Q = 2.7 \times 10^6$ to convert from normalized time to physical time, $t_{\text{phys}}$. The expresssion for the red-dashed curve is

$$F = \sqrt{1 + (\alpha - 1)^2}. \tag{10}$$

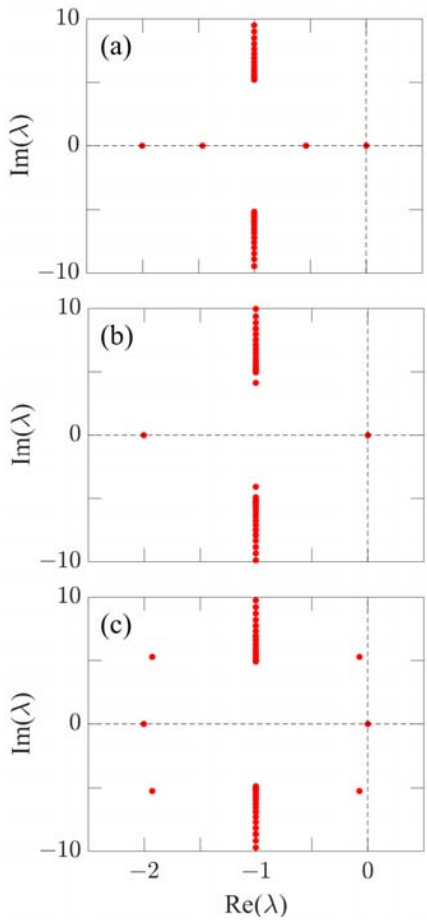

**Fig. 12.** Dynamical spectra of single solitons for $\alpha = 5.5$ and (a) $F = 2.11$, (b) $F = 2.45$, and (c) $F = 2.9$ at $L = 50$.

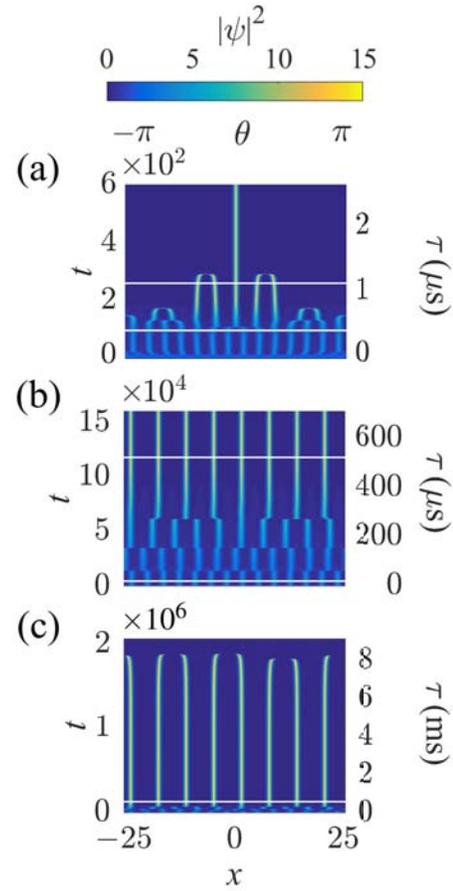

**Fig. 13.** Evolution of the waveform when the system parameters move along the red-dashed curve in Fig. 8(a). (a) The lower white line corresponds to $t = 100$ and the upper white line corresponds to $t = 250$, inside of which the detuning shifts from $\alpha = 0$ to $\alpha = 6$. At the end of the evolution, a single soliton appears. (b) The lower white line corresponds to $t = 100$ and the upper white line corresponds to $t = 1.201 \times 10^5$, inside of which the detuning again shifts from $\alpha = 0$ to $\alpha = 6$. (c) The lower white line corresponds to $t = 1.201 \times 10^5$. At the end of the time, the solution has collapsed to a continuous wave.

In Fig. 13(a), we show the evolution when the initial evolution from $\alpha = 0$ to $\alpha = 6$ occurs rapidly, analogous to the case considered in [67]. In this case, the parameters are initially fixed at $\alpha = 0$, $F = 6$ and remain there until $t = 100$ ($\tau = 0.46$ μs). By that time, a stable $N_{\text{per}} = 13$ cnoidal wave has emerged. The detuning is then increased linearly up to $\alpha = 6$, arriving at $t = 250$ ($\tau = 1.15$ μs). Finally, we wait until $t = 600$ ($\tau = 2.76$ μs). The original $N_{\text{per}} = 13$ cnoidal wave that forms at $\alpha = 0$ does not have time to evolve into anything else before the system arrives at $\alpha = 6$. Subsequently, the unstable waveform evolves into a single soliton. This behavior consistently occurs with different noise realizations. In Figs. 13(b) and 13(c), we show the evolution when the transition between $\alpha = 0$ and $\alpha = 6$ occurs between $t = 100$ and $t = 1.201 \times 10^5$ ($\tau = 0.55$ ms). We then allow the solution to evolve up to $t = 2 \times 10^6$ ($\tau = 9.2$ ms). In this case, the evolution is sufficiently slow for several transcritical bifurcations to take place as $\alpha$ increases. The $N_{\text{per}} = 8$ cnoidal wave that appears at the end of the initial evolution is only weakly unstable, i.e., $\max(\lambda_R) \simeq 6.7 \times 10^{-5}$. The solution that we show ultimately collapses to a continuous wave at around 8 ms, as shown in Fig. 13(c). We have run this simulation with four other noise realizations. In all cases, the solution collapses to continuous waves. The computational integration time is quite long in this case, although the physical time is less than a second. Again, the use of the dynamical approach to point to the presence of an instability was critical in integrating sufficiently long to detect it.

We find that the final state of the system depends on how quickly the system moves through the parameter space, as well as the trajectory.

## 5. CONCLUSION

Single solitons are a special case ($N_{\text{per}} = 1$) of cnoidal waves, more commonly referred to in the microresonator community as Turing rolls. We have determined the parameter ranges within which different periodicities are stable by solving the Lugiato-Lefever equation for parameters that are relevant for microresonators. We have also described methods for accessing these dif-ferent periodicities. We have demonstrated that cnoidal waves with $N_{\text{per}} > 1$ can have a broad bandwidth, comparable to single solitons, while at the same time they are easier to access and use the pump more efficiently. In this limit, they are effectively a periodic train of solitons or a soliton crystal.

In order to determine the ranges of stable operation in the parameter space and to optimize the cnoidal wave parameters, we used a set of software algorithms that are based on dynamical systems theory. While the basic ideas are old, we are the only research group in optical sciences of which we are aware that has implemented them in software in a form that allows us to study any stationary waveform. As part of these algorithms, we calculate the dynamical spectrum, which is the set of eigenvalues for the linearized operator about a stationary waveform. In addition to its utility in determining the stability of the stationary waveform, the dynamical spectrum also allows us to calculate the time scale on which an instability will manifest itself.

We focus in particular on the $N_{\text{per}} = 8$ cnoidal waves for a normalized mode circumference $L = 50$, which corresponds approximately to the experimental parameters of [67]. We describe the mechanisms by which the cnoidal wave can become unstable or cease to exist.

The cnoidal waves for a fixed device length occupy an approximately U-shaped band in the pump power-detuning parameter space. Below this band, only continuous waves are stable. Above this band, only breathers or chaotic solutions exist. For a normalized detuning $\alpha < 41/30$, cnoidal waves are easily accessed by simply raising the pump power. That is the case for the $N_{\text{per}} = 8$ cnoidal wave. Once a cnoidal wave has been accessed, its parameters can be changed by moving inside its stable region in the parameter space. Typically cnoidal waves with several different periodicities can exist for the same system parameters. We have found that they all have comparable bandwidths. In particular, that is the case for cnoidal waves that exist for the same set of parameters as single solitons.

The most common way to obtain single solitons is to start with negative detunings, where continuous waves or high-periodicity cnoidal waves exist. The system is then red-detuned through the high-periodicity cnoidal wave region into the chaotic region. After further red detuning, the system moves into the region where low-periodicity cnoidal waves, including single solitons, exist. This path through the chaotic region makes it possible for multiple solitons to appear, whose number and spacing vary randomly from shot to shot. Since the region where single solitons are stable nearly overlaps with the regions where continuous-waves and several other low-periodicity cnoidal waves exist, it is hard to deterministically ensure that only a single soliton will appear.

We have investigated an alternative approach in which the system moves through a U-shaped trajectory in the parameter space where cnoidal waves are stable. We showed that the stationary cnoidal wave that is obtained depends on the time scale at which the system moves along a trajectory through this space, as well as the trajectory itself.

All the theoretical work presented here is based on the Lugiato-Lefever equation. We have not discussed noise issues in any detail, and we have not discussed higher-order dispersion or thermal effects. It is reasonable to suppose that techniques that use dispersive waves to increase the bandwidth of single solitons would be useful for cnoidal waves, and could perhaps stabilize them against the effects of noise, in combination with acoustic effects, as is the case in some fiber lasers with multiple pulses in the cavity [19]. It is already known that thermal effects have a profound effect on the region in the parameter space where single solitons are stable [29, 34, 70, 71]. These issues remain to be examined.

## FUNDING INFORMATION

Funding at UMBC: DARPA/AMRDEC (W31P4Q-14-1-0002); NSF (ECCS-1807272). Funding at Purdue: NSF (ECCS-1809784); AFOSR (FA9550-15-1-0211).

## ACKNOWLEDGMENTS

We thank Yanne Chembo and Omri Gat for usefull discussions. We thank Prem Kumar for encouraging and supporting this work. A portion of our computing work was carried out at the UMBC High Performance Computing Center (https://hpcf.umbc.edu/).

## SEE SUPPLEMENT

# Dissipative cnoidal waves (Turing rolls) and the soliton limit in microring resonators: supplementary material

Zhen Qi[1,*], Shaokang Wang[1], José Jaramillo-Villegas[2], Minghao Qi[3], Andrew M. Weiner[3], Giuseppe D'Aguanno[1], Thomas F. Carruthers[1], and Curtis R. Menyuk[1]

[1]*University of Maryland at Baltimore County, 1000 Hilltop Circle, Baltimore, MD 21250, USA*
[2]*Technological University of Pereira, Cra. 27 10-02, Pereira, Risaralda 660003, Colombia*
[3]*Purdue University, 610 Purdue Mall, West Lafayette, IN 47907, USA*
*\*Corresponding author: zhenqi1@umbc.edu*

**This document provides supplementary information for the article "Dissipative cnoidal waves (Turing rolls) in microring resonators". In Sec. S1, we show stability maps with normalized detuning. In Sec. S2, we discuss some features of the dynamical spectrum of the linearized Lugiato-Lefever equation. In Sec. S3, we give our formulation of the quantum noise modeling. In Sec. S4, we give a complete stability for the cnoidal waves where $L = 50$ and $-2 < \alpha < 6$. In Sec. S5, we derive an expression for the asymptotic behavior $L/N_{\mathrm{per}}$ as $L \to \infty$.** © 2019 Optical Society of America under the terms of the OSA Open Access Publishing Agreement

***OCIS codes:*** (190.4380) Nonlinear optics, four-wave mixing; (140.4780) Optical resonators; (190.3270) Kerr effect.

http://dx.doi.org/10.1364/osa.XX.XXXXXX

## S1. STABILITY MAPS WITH NORMALIZED DETUNING

In this section, we show maps of the regions where stable wave solutions exist using the normalizations in Eq. (4). These plots are analogous to the plots in Figs. 3 and 8. In Fig. S1, we use the evolutionary method to find stationary solutions, and we integrate up to $t = 1000$. Each symbol in the figure corresponds to a different choice of the initial parameters $h$ and $\delta$. We show results for $L_\delta = 50$ and $L_\delta = 100$ with two different initial conditions. For the low-amplitude initial condition, we use an initial amplitude of $10^{-5}$ at $x = 0$, and for the high-amplitude initial condition, we use an amplitude of $10^5$ at $x = 0$. The cnoidal wave solutions occupy an approximately cone-shaped region in the $h$-$\delta$ parameter space. Which cnoidal wave solution appears depends on the initial condition, as well as the parameters. In particular, low-periodicity cnoidal waves are never found with the low-amplitude initial condition. The lowest periodicity that we observed with $L_\delta = 50$ is $N_{\mathrm{per}} = 7$, and the lowest periodicity that we observed with $L_\delta = 100$ is $N_{\mathrm{per}} = 15$. By contrast, low-periodicity cnoidal waves, including solitons, can be obtained by starting with the high-amplitude initial condition. As was the case with the normalization of Eq. (3), the location of the stability boundaries is ambiguous.

In Fig. S2, we show a map of the stable regions for $L_\delta = 50$ and $L_\delta = 100$ that we obtained using the dynamical approach. The self-similarity that we previously described in Sec. III is apparent. The stable regions for $N_{\mathrm{per}} = X$ at $L_\delta = 50$ are approximately the same as the stable regions for $N_{\mathrm{per}} = 2X$ at $L_\delta = 100$. As the periodicity increases, the loss that is necessary to obtain stable cnoidal waves also inreases. Hence, these stable solutions are disconnected from the lossless analytical solutions that we previously obtained [S1], which limits their utility.

## S2. DYNAMICAL SPECTRUM OF THE LINEARIZED LLE

In this section, we will derive some properties of the dynamical spectrum of the linearized LLE.

It is useful to first transform $\Delta\mathbf{\Psi}$, given in Eqs. (6)–(8) to remove the attenuation by defining

$$\Delta\mathbf{X} = \begin{bmatrix} \Delta\chi \\ \Delta\bar{\chi} \end{bmatrix} = \Delta\mathbf{\Psi}\exp t = \begin{bmatrix} \Delta\psi\exp t \\ \Delta\bar{\psi}\exp t \end{bmatrix}. \quad (S.1)$$

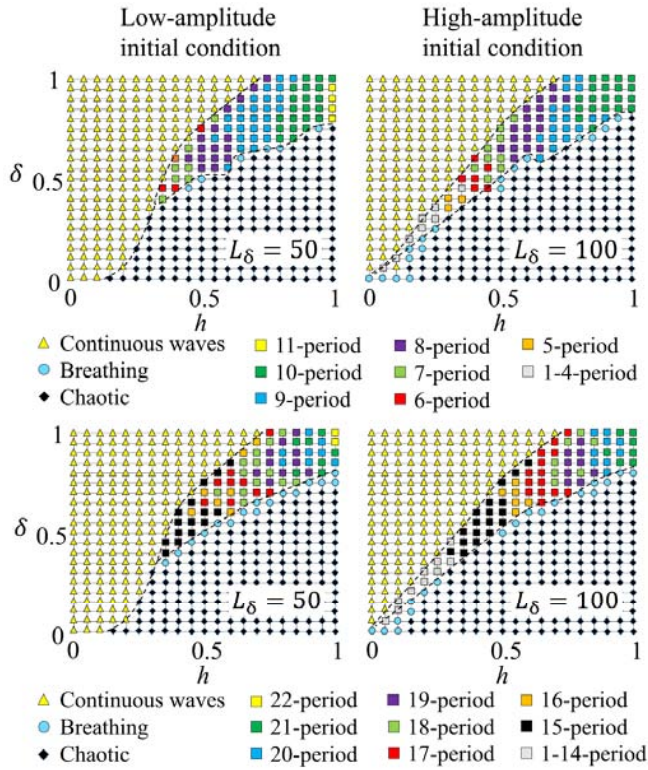

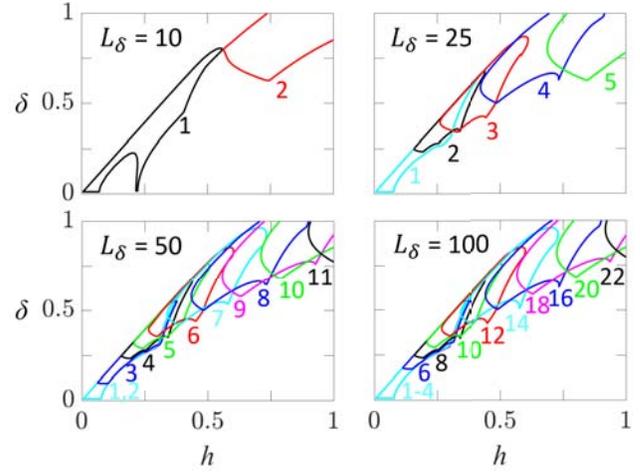

**Fig. S1.** We show a map of the solutions that emerge using the evolutionary approach for $L_\delta = 50$ and 100 and for two different initial conditions.

**Fig. S2.** Maps of the stable regions for cnoidal waves for $L_\delta = 10$, 25, 50 and 100.

This transformation shifts all the eigenvalues by +1 so that the fourfold symmetry now appears with respect to the real and imaginary axes. We note that we have made no change in $\psi_0$, which has embedded in it the effects of both the pump and attenuation. We next let $\psi_0 = r_0 + is_0$, $\Delta\chi = r + is$, and $\Delta\bar{\chi} = r - is$. We also let $\mathbf{z}(t) = [r(t), s(t)]^T$, where we use $T$ to denote the transpose so that $\mathbf{z}$ is a column vector. We now find

$$\frac{\partial \mathbf{z}}{\partial t} = \frac{\partial}{\partial t}\begin{bmatrix} r \\ s \end{bmatrix} = \begin{bmatrix} \mathbf{M}_{11} & \mathbf{M}_{12} \\ \mathbf{M}_{21} & \mathbf{M}_{22} \end{bmatrix} = \mathbf{M}\mathbf{z}, \quad (S.2)$$

where

$$\mathbf{M}_{11} = -2r_0 s_0, \quad \mathbf{M}_{12} = -\frac{\partial^2}{\partial x^2} - r_0^2 - 3s_0^2 + \alpha,$$

$$\mathbf{M}_{21} = \frac{\partial^2}{\partial x^2} + 3r_0^2 + s_0^2 - \alpha, \quad \mathbf{M}_{22} = 2r_0 s_0. \quad (S.3)$$

The operator $\mathbf{M}$ is real, which is sufficient to imply that if $\lambda$ is a non-real eigenvalue, then $\lambda^*$ is also an eigenvalue. Hence, the eigenvalues are symmetric with respect to the real axis. This same requirement holds for the linearized equation that corresponds to any variant of the scalar nonlinear Schrödinger equation, not just the LLE [S2].

The key to understanding the symmetry with respect to the imaginary axis is that the evolution equations that govern $\mathbf{z}(t)$ are Hamiltonian. These equations are derivable from the Hamiltonian

$$\mathcal{H} = \int_{-L/2}^{L/2} dx \left[ \frac{1}{2}\left(\frac{\partial r}{\partial x}\right)^2 + \frac{1}{2}\left(\frac{\partial s}{\partial x}\right)^2 - \left(r_0^2 + s_0^2 - \frac{\alpha}{2}\right)(r^2 + s^2) \right. $$
$$\left. - \frac{1}{2}(r_0^2 - s_0^2)(r^2 - s^2) - 2r_0 s_0 rs \right]. \quad (S.4)$$

We then find $\partial r/\partial t = D\mathcal{H}/Ds$ and $\partial s/\partial t = -D\mathcal{H}/Dr$, where we use D to denote the functional derivative. Hence, we find that $r(x,t)$ is a canonical coordinate, parameterized by $x$, and that $s(x,t)$ is the corresponding canonical momentum. We may also write the evolution equations as

$$\frac{\partial \mathbf{z}}{\partial t} = \mathbf{M}z = \mathbf{J}\frac{D\mathcal{H}}{D\mathbf{z}}, \quad (S.5)$$

where $\mathbf{J}$ is the symplectic operator. It is defined by

$$\mathbf{J}\begin{bmatrix} a(x) \\ b(x) \end{bmatrix} = \begin{bmatrix} b(x) \\ -a(x) \end{bmatrix}, \quad (S.6)$$

where $a$ and $b$ are arbitrary functions of $x$. The operator $\mathbf{M}$ shares with all real Hamiltonian operators for linear systems the properties $\mathbf{M}_{22} = -\mathbf{M}_{11}^T$, $\mathbf{M}_{12} = \mathbf{M}_{12}^T$, and $\mathbf{M}_{21} = \mathbf{M}_{21}^T$ [S3]. We now find

$$\mathbf{J}(\mathbf{M} - \lambda \mathbf{I})\mathbf{J} = \mathbf{JMJ} - \lambda \mathbf{J}^2 = \mathbf{M}^T + \lambda \mathbf{I}. \quad (S.7)$$

Since the eigenvalues of $\mathbf{M}$ and $\mathbf{M}^T$ are the same, we conclude that if $\lambda$ is an eigenvalue, then so is $-\lambda$ and hence so is $\lambda^*$. So, the dynamical spectrum is symmetric about the imaginary axis.

We note parenthetically that this Hamiltonian formulation is a useful starting point for investigating quantum effects in microresonators.

We found that a number of the eigenvalues associated with the $N_{\text{per}} = 8$ cnoidal waves are degenerate. Similar degeneracies occur with other cnoidal waves. Degeneracy is a consequence of translational symmetry. The length of one period of a cnoidal wave is $L/N_{\text{per}}$. It follows that if $\Delta\mathbf{\Psi}_{\lambda 1}(x)$ is an eigenvector with an eigenvalue $\lambda$, then so is $\Delta\mathbf{\Psi}_{\lambda 1}(x+L/N_{\text{per}})$. We must be able to write

$$\Delta\mathbf{\Psi}_{\lambda 1}(x+L/N_{\text{per}}) = \sum_{m=1}^{M} c_m \Delta\mathbf{\Psi}_{\lambda m}(x), \quad (S.8)$$

where the $c_m$ are constants, the $\Delta\mathbf{\Psi}_{\lambda m}(x)$ are the eigenvectors that share the same eigenvalue $\lambda$, and $M$ is the total number of eigenvectors that share that eigenvalue. If an eigenvalue is simple, then its eigenvector must have a period equal to $L/N_{\text{per}}$, which will not be the case in general. Thus, a high degree of degeneracy is expected for cnoidal waves with large $N_{\text{per}}$.

## S3. QUANTUM NOISE SIMULATION

The fluctuating energy due to quantum noise is at least $h\nu$ per mode, where $h$ is Planck's constant and $\nu$ is the central frequency of the mode. In a system with damping, nature must supply sufficient noise power to guarantee this noise level. The modes in a microresonator are spaced apart in frequency by the free spectral range (FSR) $1/T_R$, where $T_R$ is the round-trip time. In practice, this fluctuating energy limit is just a lower limit. The actual fluctuating energy will typically be higher due to electromechanical and environmental noise sources [S4].

Writing $U_m = h\nu$ for the $m$-th mode, we obtain $P_m = h\nu/T_R$, where $P_m$ is the corresponding power in the microresonator. It might seem strange that the noise power in mode $m$ increases as the frequency separation increases. This increase occurs because the device size shrinks as the FSR increases, so that the energy density of each mode increases, and the amount of energy passing through any point of the microresonator per unit time increases.

The amplitude of the $m$-th mode is given by the Fourier transform of $A(\tau, \theta)$,

$$\tilde{A}_m(\tau) = \int_0^{2\pi} \frac{d\theta}{2\pi} A(\tau, \theta) \exp(im\theta). \quad (S.9)$$

The number of different modes that are kept in a simulation $N_{\text{mode}}$ is equal to the number of node points in the simulation. We thus find that

$$\tilde{A}_m(\tau) = \frac{1}{N_{\text{mode}}} \sum_{j=1}^{N_{\text{mode}}} A(\tau, \theta_j) \exp(im\theta_j) \quad (S.10)$$

after discretization, where $\theta_j = 2\pi j/N_{\text{mode}}$. The average noise power due to quantum fluctuations is given by $P_m = \langle|\tilde{A}_{\text{noise},m}|^2\rangle = h\nu/T_R$, where the brackets $\langle\cdot\rangle$ indicate an average over $\tau$. We then find that the average noise power due to quantum fluctuations at each $\theta_j$ is given by $\langle|A_{\text{noise}}(\theta_j)|^2\rangle = (h\nu/T_R)N_{\text{mode}}$. The loss due to attenuation must be compensated by vacuum fluctuations. In a system with just attenuation, we find [S5],

$$T_R \frac{d\tilde{A}_{\text{noise},m}}{d\tau} = -\frac{l}{2}\tilde{A}_{\text{noise},m} + R_m, \quad (S.11)$$

where $\langle R_m(\tau) R_{m'}^*(\tau')\rangle = h\nu l \delta(\tau-\tau')\delta_{m,m'}$. It follows that the average change in $\tilde{A}_{\text{noise},m}$ in a small time step $\Delta\tau$ due to noise is given by

$$\begin{aligned}\langle|\Delta\tilde{A}_{\text{noise},m}|^2\rangle &= (h\nu/T_R)\left[1-\exp(-l\Delta\tau/T_R)\right] \\ &\simeq (h\nu/T_R)(l\Delta\tau/T_R).\end{aligned} \quad (S.12)$$

The corresponding change at each node point $\theta_j$ is given by $\langle|\Delta A(\theta_j)|^2\rangle = (h\nu/T_R)(l\Delta\tau/T_R)N_{\text{mode}}$. The key point is that vacuum fluctuations do not change in the presence of dispersion and an external pump that compensates for loss [S5]. Using the transformations following Eq. (1), we find that the corresponding changes in $\Delta\tilde{\psi}_m$ and $\Delta\psi(x_j)$ over a small time step $\Delta t$ are given by $\langle|\Delta\tilde{\psi}_m|^2\rangle = (2\gamma/l)(h\nu/T_R)2\Delta t$ and $\langle|\Delta\psi(x_j)|^2\rangle = (2\gamma/l)(h\nu/T_R)2\Delta t N_{\text{mode}}$.

The computational algorithm is to add Gaussian-distributed random noise to the real and imaginary part of $\psi$ at each node point $\theta_j$, whose variance for each separately is equal to $(1/2)(2\gamma/l)(h\nu/T_R)2\Delta t N_{\text{mode}}$. Alternatively, one can add the noise in the wavenumber domain, using the variances for the real and imaginary parts of $\Delta\tilde{\psi}_m$.

For the parameters of Jaramallo-Villegas, et al. [S6] and assuming a step size that is $0.01 T_R$, we first find that $l\Delta\tau/T_R = 2\Delta t = 0.02$. The noise power that must be added on each time is given by

$$\begin{aligned}\langle|\Delta\tilde{A}_m|^2\rangle &= (h\nu/T_R)(l\Delta\tau/T_R) \\ &= (6.63\times 10^{-34})(2.0\times 10^{14})(2.26\times 10^{11})(0.02) \\ &= 5.99\times 10^{-10}\text{ W}.\end{aligned}$$
(S.13)

If we assume that there are 512 nodes, which is a typical value in simulations, then we find that $\langle|\Delta A(\theta_j)|^2\rangle = 3.07\times 10^{-7}$ W. This power corresponds to approximately $-35$ dBm, which is small, but larger by about 200 dBm from the noise power due to roundoff noise in typical simulations. The corresponding value of $\langle|\Delta\tilde{\psi}_m|^2\rangle$ is $6.03\times 10^{-18}$, and the corresponding value of $\langle|\Delta\psi(x_j)|^2\rangle$ is $3.09\times 10^{-15}$.

## S4. COMPLETE MAP OF THE STABLE REGIONS FOR $L=50$

In Fig. S3, we show a map of the stable regions for all the cnoidal waves that are stable in the range $-2<\alpha<6$. This map is presented as a slide show in which one of the stable regions is highlighted in each slide.

## S5. ASYMPTOTIC VALUE OF $L/N_{\text{per}}$ AS $L\to\infty$

Here we calculate the most unstable wavenumber of continuous waves for parameters where cnoidal waves are stable and continuous waves are unstable. That allows us to predict analytically which cnoidal wave will form starting from noise when $L\to\infty$ and to obtain the asymptotic value of $L/N_{\text{per}}$. This procedure is similar to the one used by Godey et al. [S7] to predict which cnoidal wave will appear when continuous waves go unstable.

We start by assuming that $\psi_0$ is a complex constant. We then find that $\rho = |\psi_0|^2$ is given by the solution to the cubic equation

$$[1+(\alpha-\rho)^2]\rho = F^2. \quad (S.14)$$

We now perturb this solution, using the ansatz $\Delta\psi = \exp(\lambda t + ikx)$, $\Delta\bar{\psi} = \Delta\psi^*$. Substitution into Eq. (7) yields $\lambda = -1\pm[\rho^2-(2\rho-\alpha-k^2)^2]^{1/2}$. The growth rate $\lambda$ is maximized when $dk/d\lambda = 0$, which implies $k = (2\rho-\alpha)^{1/2}$ or

$$L/N_{\text{per}} = 2\pi/k = \frac{2\pi}{(2\rho-\alpha)^{1/2}}. \quad (S.15)$$

Substitution of the solution of Eq. (S.14) into Eq. (7) permits us to compute $L/N_{\text{per}}$. For $\alpha = -2$, $F = 3.5$, we find $L/N_{\text{per}} = 3.04$; for $\alpha = -1$, $F = 2.6$, we find $L/N_{\text{per}} = 3.43$; for $\alpha = 0$ and $F = 1.7$, we find $L/N_{\text{per}} = 4.07$; for $\alpha = 1$, $F = 1.2$, we find $L/N_{\text{per}} = 4.93$.

**Fig. S3.** A slide show map of all the stable regions for the cnoidal waves at $L = 50$ in the range $-2 < a < 6$. One of the stable regions is highlighted in each slide; the regions are labeled with their corresponding periodicity $N_{\text{per}}$. The red-dashed line shows the boundary below which continuous waves are stable.